\documentclass[preprint,12pt]{elsarticle}

\usepackage{amssymb}
\usepackage{amsmath}
\usepackage{booktabs}
\usepackage{subcaption}
\journal{Journal of Applied Geophysics}

\begin{document}

\begin{frontmatter}




\title{Seismic first-arrival traveltime simulation based on reciprocity-constrained PINN}


\author[1]{Hang Geng}
\author[1]{Chao Song}
\author[2]{Umair bin Waheed}
\author[1]{Cai Liu}

\affiliation[1]{organization={College of Geo-Exploration Science and Technology,Jilin University},
            addressline={Ximinzhu},
            city={Changchun},
            postcode={130026},
            state={Jilin Province},
            country={China}}

\affiliation[2]{organization={King Fahd University of Petroleum and Minerals (KFUPM)},
            city={Dhahran},
            postcode={31261},
            country={Saudi Arabia}}

\begin{abstract}
Simulating seismic first-arrival traveltime plays a crucial role in seismic tomography. First-arrival traveltime simulation relies on solving the eikonal equation. The accuracy of conventional numerical solvers is limited to a finite-difference approximation. In recent years, physics-informed neural networks (PINNs) have been applied to achieve this task. However, traditional PINNs encounter challenges in accurately solving the eikonal equation, especially in cases where the model exhibits directional scaling differences. These challenges result in substantial traveltime prediction errors when the traveling distance is long. To improve the accuracy of PINN in traveltime prediction, we incorporate the reciprocity principle as a constraint into the PINN training framework. Based on the reciprocity principle, which states that the traveltime between two points remains invariant when their roles as source and receiver are exchanged, we propose to apply this principle to multiple source-receiver pairs in PINN-based traveltime prediction. Furthermore, a dynamic weighting mechanism is proposed to balance the contributions of the eikonal equation loss and the reciprocity-constrained loss during the training process. This adaptive weighting evolves dynamically with the training epochs, enhancing the convergency of the training process. Experiments conducted on a simple lens velocity model, the Overthrust velocity model, and a 3D velocity model demonstrate that the introduction of the reciprocity-constrained PINN significantly improves the accuracy of traveltime predictions.
\end{abstract}


\begin{highlights}
\item The effective-slowness factored eikonal equation is utilized, eliminating source-velocity dependency and avoiding extra parameter tuning.
\item A reciprocity-constrained PINN (RC-PINN) is proposed to enhance traveltime prediction accuracy.
\item A dynamic weighting strategy adaptively balances the eikonal equation loss and reciprocity-constrained loss, improving training convergence without additional computational cost.
\item Result on 2D lens, Overthrust, and 3D velocity models demonstrates the superiority of RC-PINN.
\end{highlights}


\end{frontmatter}


\section{Introduction}\label{sec1}

Seismic tomography is a foundational technology in geophysics and seismic exploration \cite{bib1}. By analyzing variations in seismic first-arrival traveltime misfit between the observed and predicted seismic waves \cite{bib3,bib4}, it reconstructs the Earth's internal velocity structure, revealing the physical properties of subsurface medium \cite{bib2}. This technology is instrumental not only in oil and gas but also in earthquake disaster prediction \cite{bib5}, volcanic activity monitoring, and the study of deep Earth structures. In exploration seismology, seismic tomography is pivotal for developing accurate velocity models, which are indispensable for the application of high-precision seismic imaging techniques, such as reverse time migration (RTM) \cite{bib6} and full waveform inversion (FWI) \cite{bib7,bib8,bib9}.

The high-frequency approximation ray theory imaging method \cite{bib10}, the finite-frequency imaging method \cite{bib11}, and the traveltime tomography method are popularly used traveltime tomography methods \cite{bib12}. Notably, the tomography method utilizing the eikonal equation offers significant advantages. By directly inferring velocity gradients through the optimization of differences between simulated and observed traveltime, this method enhances the stability of the velocity inversion process \cite{bib13,bib14}.

In first-arrival traveltime tomography, the accuracy of first-arrival traveltime simulation is a critical step in the inversion process \cite{bib16}. The eikonal equation is a widely used tool for seismic wave first-arrival traveltime simulation \cite{bib17}. This non-linear partial differential equation effectively describes seismic wave propagation paths and arrival times across various media.

Traditional numerical methods for solving the eikonal equation, like the fast marching method (FMM) and the fast sweeping method (FSM) are popularly used \cite{bib18,bib19}. These methods simulate wavefront propagation on regular grids, making them well-suited for first-arrival traveltime computations. However, as models grow more complex, such as in scenarios involving complex geological structures or high-contrast velocity variations, these methods face limitations in both accuracy and efficiency. Additionally, in irregular grids or heterogeneous media, finite-difference approaches may suffer from cumulative errors, leading to deviations in first-arrival traveltime results.

The emergence of physics-informed neural networks (PINNs) has recently introduced a novel perspective in this field \cite{bib20,bib21}. By embedding physical laws into neural network training as constraints, PINNs allow the networks to learn physical principles. Applications of PINNs to time-domain and frequency-domain seismic wavefield simulations and inversions across diverse media have demonstrated their impressive effectiveness \cite{bib22,bib23,bib24,bib25,bib26,bib27}.

In solving the eikonal equation, PINNs provide a promising alternative for managing complex geometries that challenge traditional numerical methods \cite{bib28,bib29}. For instance, in irregular grids or heterogeneous media, where finite-difference methods often accumulate errors, PINNs leverage automatic differentiation to evaluate partial derivatives accurately and mitigate these challenges caused by the finite-difference approximation, resulting in improved accuracy for first-arrival traveltime simulation. Moreover, due to the flexible architecture of the networks, we can use source locations as input to achieve multi-source traveltime simulation.

Despite the advantages, PINNs also face challenges. PINNs accept the vertical and horizontal model coordinates as input. When PINNs are applied to velocity models with significant directional scale differences, such as large lateral-to-depth ratios, the values of input horizontal and vertical coordinates are not balanced, resulting in increasing traveltime errors as the travel distance gets larger.

In this paper, we propose a novel method named reciprocity-constrained PINN (RC-PINN) to improve the performance of PINN-based eikonal solvers by imposing the reciprocity principle as a constraint. By selecting multiple source and receiver pairs, this approach minimizes their traveltime misfits to honor the reciprocity principle. Additionally, a dynamic weighting strategy is introduced between the physics-constrained loss (eikonal equation) and the reciprocity-constrained loss, enhancing the training convergence and traveltime prediction accuracy.

\section{Theory}\label{sec2}
\subsection{Eikonal equation and its effective-slowness factored form}\label{subsec2}

The first-arrivel
traveltime of seismic waves can be simulated by solving the eikonal equation given by \cite{bib30}\cite{bib31}:
\begin{equation}
|\nabla T(\mathbf{x})|^2 = \frac{1}{v^2(\mathbf{x})}, 
\end{equation}
\begin{equation}
T(\mathbf{x}_s) = 0,
\end{equation}
where \(T(\mathbf{x})\) is the traveltime at position \(\mathbf{x}\), \(v(\mathbf{x})\) is the velocity of the medium at position \(\mathbf{x}\), \(\nabla T(\mathbf{x})\) is the gradient of \(T(\mathbf{x})\), and \(\mathbf{x}_s\) is the location of the source.

The source singularity issue can cause inaccurate traveltime solutions near the source location. To address issue, the factored form of the eikonal equation can be employed by decomposing \(T(\mathbf{x})\) as follows \cite{bib32}:
\begin{equation}
T(\mathbf{x}) = T_{0}(\mathbf{x})\gamma(\mathbf{x}),
\end{equation}
where \(T_{0}(\mathbf{x})\) denotes the background traveltime term from a constant wave velocity value, \(T_{0}(\mathbf{x})\) can be easily calculated as:
\begin{equation}
T_0(\mathbf{x}) = \frac{|\mathbf{x} - \mathbf{x}_s|}{v_0}, 
\end{equation}
where \( |\mathbf{x} - \mathbf{x}_s| \) represents the distance from \( \mathbf{x} \) to the source point \( \mathbf{x}_s \), and \( v_0 \) denotes the velocity at the source location. \( \gamma \) represents the traveltime factor, which can be expressed as:
\begin{equation}
\gamma = \frac{T}{T_0} = \frac{\frac{|\mathbf{x} - \mathbf{x}_s|}{v'}}{\frac{|\mathbf{x} - \mathbf{x}_s|}{v_0}} = \frac{v_0}{v'},
\end{equation}
where \( v' \) is the effective velocity between \( \mathbf{x} \) and the source \( \mathbf{x}_s \). Since \( v' \) falls within the range \([v_{\min}, v_{\max}]\), where \( v_{\max} \) and \( v_{\min} \) represent the maximum and minimum velocities of the velocity model, respectively. The range of \(\gamma\) falls within 
\(\left[\frac{v_{0}}{v_{\max}}, \frac{v_{0}}{v_{\min}}\right]\),It can be seen that the range of \( \gamma \) varies with the change in the velocity at the source location for different sources. This makes \( \gamma \) difficult to predict, leading to challenges in prediction during the training process. Therefore, we adopted a new decomposition form, which is shown as follows \cite{bib33}:
\begin{equation}
T(\mathbf{x}) = R(\mathbf{x})\tau(\mathbf{x}), 
\end{equation}
where \(R(\mathbf{x})\) is the distance function, defined as:
\begin{equation}
R(\mathbf{x}) = |\mathbf{x} - \mathbf{x}_s|, 
\end{equation}
where \(\tau\) has units of \(\text{s/km}\) and represents the new traveltime factor, which can be expressed as:
\begin{equation}
\tau = \frac{T}{R} = \frac{\frac{|\mathbf{x} - \mathbf{x}_s|}{v'}}{|\mathbf{x} - \mathbf{x}_s|} = \frac{1}{v'},
\end{equation}
respectively.The range of \(\tau\) is 
\(\left[\frac{1}{v_{\max}}, \frac{1}{v_{\min}}\right]\). Compared to \(\gamma\), The range of \(\tau\) under different sources is fixed and does not change with the velocity at the source location. This makes traing convergence of PINN easier, thereby improving training efficiency. Moreover, in most cases, \(v_{\min} > 1\), ensuring that \(\tau\) falls within the value range of the sigmoid activation function, which avoids an extra parameter tuning step.

\subsection{Physics-informed neural network}\label{subsec2}

\begin{figure}[htbp]
    \centering
    \includegraphics[width=0.5\textwidth]{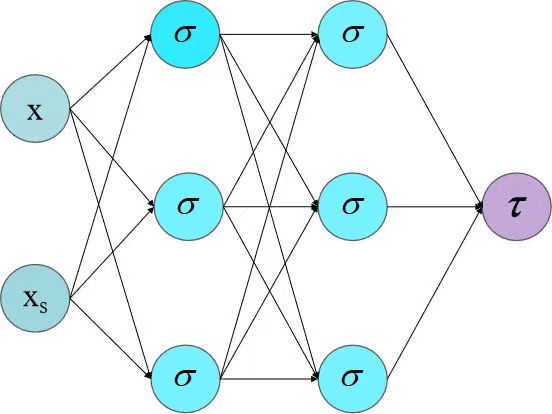}
    \caption{\centering Traveltime prediction neural network.}
    \label{netrual}
\end{figure}

In PINN-based eikonal equation solver, we can predict the first-arrival traveltime by minimizing the residuals of the factored eikonal equation. According to the universal
approximation theorem of neural networks, we use a neural network to approximate the traveltime factor $\tau$, and its network structure is shown in Figure~\ref{netrual}. We take the spatial coordinate \( \mathbf{x} \)  and the source coordinate \( \mathbf{x}_s \) as input to the neural network. The predicted traveltime can be obtained by multiplying the output of the network $\tau$ with the distance function $R$. The loss function of the PINN-based eikonal equation solver is given as:
\begin{equation}
\xi = \frac{1}{N} \sum_{i=1}^N \|\ell_1\|_2^2, 
\end{equation}
where
\begin{equation}
\ell_1 = R^2 (\nabla\tau)^2 + \tau^2(\nabla R)^2 + 2R\tau (\nabla R \nabla \tau) - \frac{1}{v^2(\mathbf{x})},
\end{equation}
where $N$ is the number of the input coordinates randomly selected within the model
domain and \(\nabla\) represents the gradient operator. 

\subsection{Reciprocity-constrained PINN}\label{subsec2}

According to the reciprocity principle, two points can be interchangeably used as the source and receiver and their traveltime should be identical. Based on this principle, for traveltime calculation for multiple sources, the traveltime \(T(\mathbf{x}_r, \mathbf{x}_s)\) from source \(\mathbf{x}_s\) to receiver \(\mathbf{x}_r\) is equal to the traveltime \(T(\mathbf{x}_s, \mathbf{x}_r)\) from receiver \(\mathbf{x}_r\) to source \(\mathbf{x}_s\), given by:
\begin{equation}
T(\mathbf{x}_r, \mathbf{x}_s) = T(\mathbf{x}_s, \mathbf{x}_r). 
\end{equation}

We apply this principle to PINN-based traveltime calculation. In this paper, we incorporate this principle as a reciprocity-constrained term alongside the physics term constrained by the factored eikonal equation. The loss function of the reciprocity-constrained PINN for traveltime prediction is expressed as:
\begin{equation}
\xi = \frac{1}{N} \sum_{i}^{N} \|\ell_1\|_2^2 + \frac{1}{2N_1} \sum_{j}^{2N_1}  \|\ell_2\|_2^2,
\end{equation}
where
\begin{equation}
\ell_2 = T\left(\mathbf{x}_r^{(j)}, \mathbf{x}_s^{(j)}\right) - T\left(\mathbf{x}_s^{(j)}, \mathbf{x}_r^{(j)}\right), 
\end{equation}
\begin{equation}
T\left(\mathbf{x}_r^{(j)}, \mathbf{x}_s^{(j)}\right) = R\left(\mathbf{x}_r^{(j)}, \mathbf{x}_s^{(j)}\right)\tau\left(\mathbf{x}_r^{(j)}, \mathbf{x}_s^{(j)}\right), 
\end{equation}
\begin{equation}
T\left(\mathbf{x}_s^{(j)}, \mathbf{x}_r^{(j)}\right) = R\left(\mathbf{x}_s^{(j)}, \mathbf{x}_r^{(j)}\right)\tau\left(\mathbf{x}_s^{(j)}, \mathbf{x}_r^{(j)}\right),
\end{equation}

where \( N_1 \) represents the number of source and receiver pairs.

\begin{figure}[htbp]
    \centering
    \includegraphics[width=0.6\textwidth]{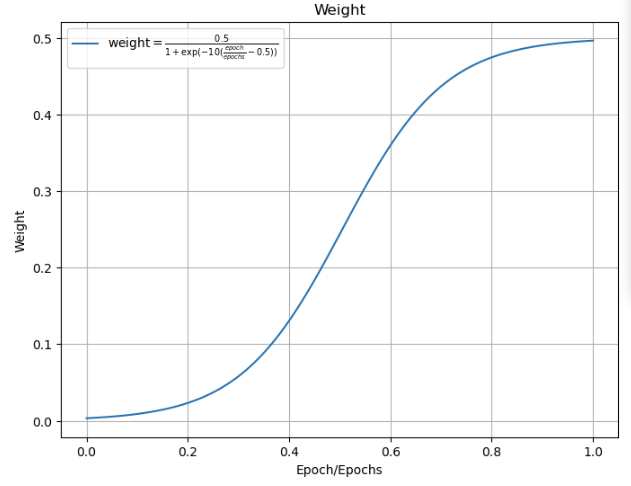}
    \caption{\centering Dynamic weight function.}
    \label{fig:dynamic_weight}
\end{figure}

To balance two loss terms in equation 12, we introduce dynamic weights that vary with training epochs. This adjusts the weight of the physics-constrained loss function to be larger at the early stage of the training, honoring the physics that traveltimes adhere. The weight of the reciprocity-constrained loss gradually increase to enhance the reciprocity principle of seismic waves. The new total loss function is defined as:
\begin{equation}
\xi = (1 - \lambda)\frac{1}{N} \sum_{i}^{N} \|\ell_1\|_2^2 + \frac{\lambda}{2N_1} \sum_{j}^{2N_1}  \|\ell_2\|_2^2, 
\end{equation}
where
\begin{equation}
\lambda = \frac{0.5}{1 + e^{-10(i / M - 0.5)}}, 
\end{equation}
\(i\) is the current training epoch and \(M\) is the number of total training epochs. The dynamic weight function is shown in Figure~\ref{fig:dynamic_weight}.

\section{Results}\label{sec2}
To demonstrate the effectiveness of the proposed RC-PINN method, we first perform a test on a constant gradient velocity model. The results obtained from the numerical algorithm, the traditional PINN method, and the RC-PINN method are compared to demonstrate the superior accuracy of RC-PINN. Subsequently, the method's performance is validated on three additional models: a simple lens velocity model, an Overthrust model, and a 3D model.
\subsection{Constant gradient velocity model}\label{subsec2}
\begin{figure}[htbp]
    \centering
    \begin{minipage}[b]{0.45\textwidth}
        \centering
        \includegraphics[width=\textwidth]{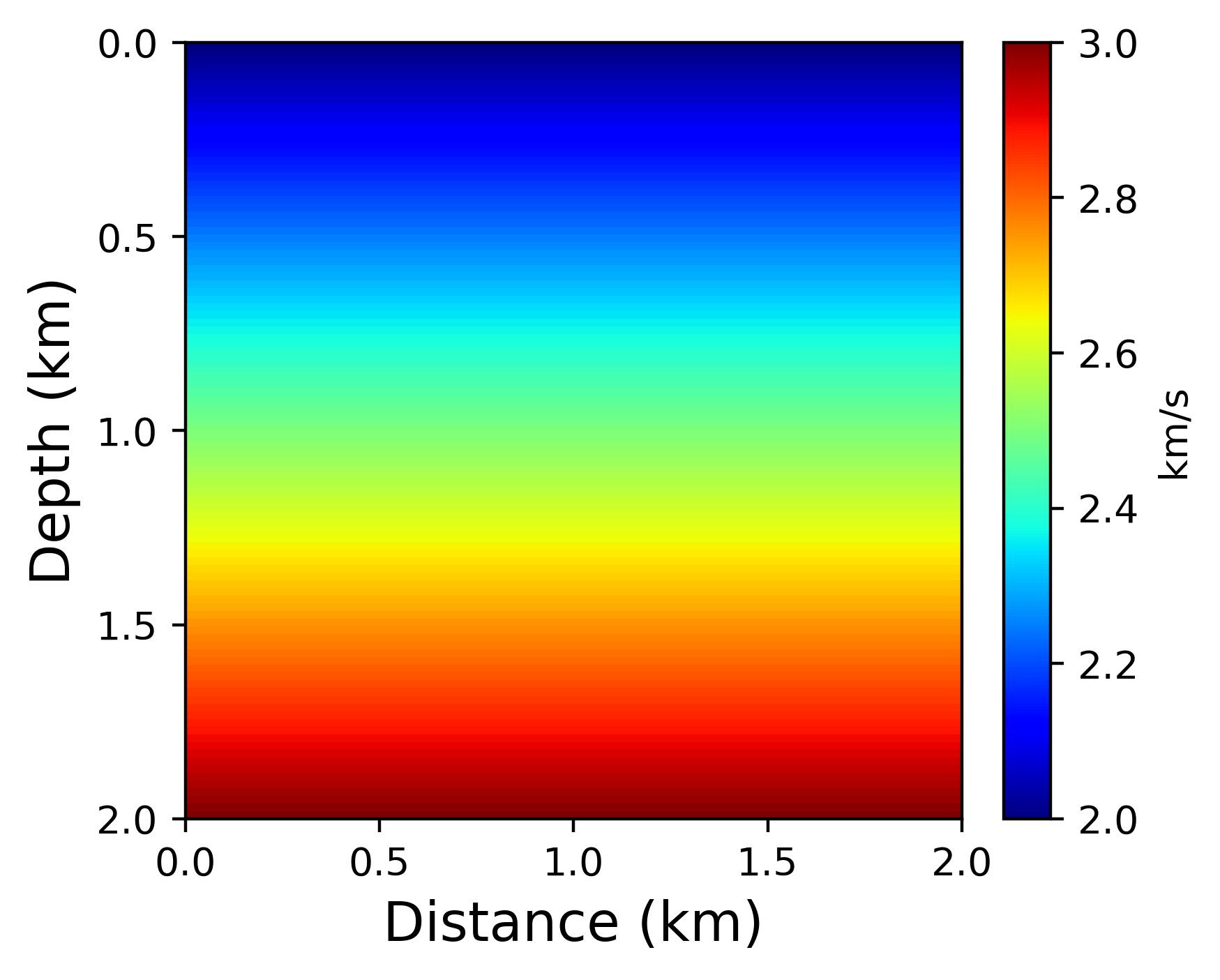}
        \caption{\centering Constant gradient velocity model. }
        \label{constant v}
    \end{minipage}
    \hfill
    \begin{minipage}[b]{0.45\textwidth}
        \centering
        \includegraphics[width=\textwidth]{	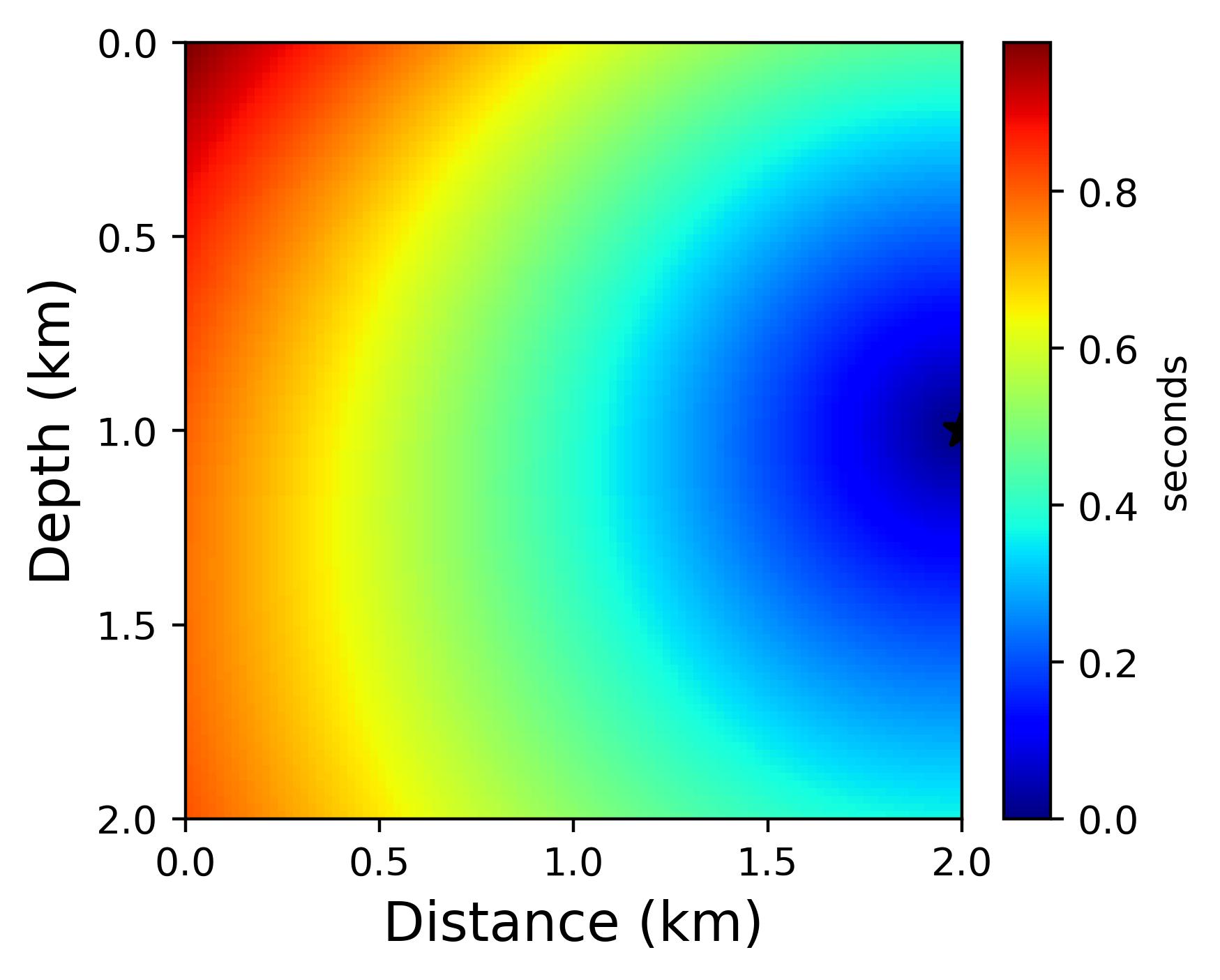}
        \caption{\centering True traveltime from analytical solution for the constant gradient velocity.}
        \label{constant t}
    \end{minipage}
\end{figure}

We begin the test with a constant gradient velocity model for traveltime simulation. The model spans a \(101 \times 101\) grid points spaced with 0.02 km apart in both horizontal and vertical directions, covering a survey area of \(2 \times 2 \,  \ \text{km}^2\). The velocity increases constantly from top to bottom, with a minimum velocity of \(2 \, \text{km/s}\) and a maximum velocity of \(3 \, \text{km/s}\). The velocity is illustrated in shown Figure~\ref{constant v}. 

The true traveltime of one source located at (1 km, 2 km) obtained from the analytical solution is illustrated in Figure~\ref{constant t}.


We configure the neural network with 6 hidden layers, each layer containing 64 neurons. A locally adaptive exponential linear unit (l-ELU) as the activation function is applied to the hidden layers, while the output layer uses a sigmoid activation function. The network is trained using the Adam full-batch optimizer. A total number of model 2000 sets of model coordinates and source coordinates are randomly generated as inputs to the neural network. The number of source and receiver pairs is 190. The traditional PINN and RC-PINN approaches are trained with their respective loss functions for 10,000 epochs. We compare the FMM-calculated, PINN-predicted, and RC-predicted traveltimes with the true traveltime from the analytical solution. Their errors are shown in Figure 5a-5c, respectively.  
\begin{figure}
    \centering
    \includegraphics[width=1\linewidth]{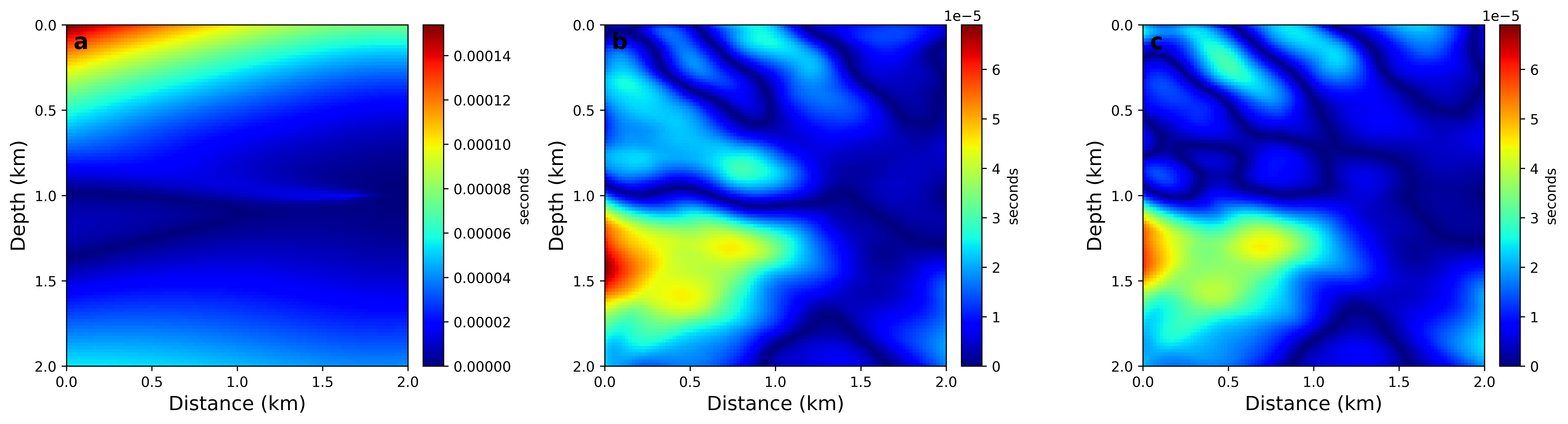}
    \caption{\centering The traveltime errors of FMM (a), traditional PINN (b) and RC-PINN (c) methods for the constant gradient velocity model. }
    \label{u_t_e}
\end{figure}

The figure illustrates that the traveltime error of FMM is relatively larger in shallow regions, particularly at greater distances, where the error exhibits a notable increase. The traveltime error of the traditional PINN also shows errors with smaller values indicating PINN's superior accuracy over FMM. In contrast, the RC-PINN method further reduces the error as shown in Figure 5c.

To provide a more quantitative analysis of the results, we quantitatively evaluate the maximum traveltime error and the L2 error for the three methods. The results are summarized in Table~\ref{1}. As shown in the table, the L2 error and the maximum error of RC-PINN are even smaller than those from the conventional PINN, which demonstrates RC-PINN's accuracy improvement from the conventional PINN.

\begin{table}[htbp]
\centering
\caption{L2 error and maximum error of traveltimes from FMM, PINN, and RC-PINN of constant gradient velocity model }\label{tab1}%
\begin{tabular}{@{}llll@{}}
\toprule
Method   & L2  Error & Maximum Error (s) \\
\midrule
FMM      & 9.12e-5            &1.52e-4             \\
PINN     & 3.81e-5            &6.72e-5             \\
RC-PINN & 3.12e-5            &5.82e-5             \\
\bottomrule
\label{1}
\end{tabular}
\end{table}

\subsection{Lens velocity model}\label{subsec2}

Next, we consider a simple lens velocity model for traveltime simulation. The model has dimensions of \(51 \times 301\),  spaced \(0.02 \, \ \text{km}\) apart in both horizontal and vertical directions, covering a \(1 \times 6 \,  \ \text{km}^2\) region. The velocity increases gradually from \(2 \, \  \text{km/s}\) at the top to \(4 \, \text{km/s}\) at the bottom, with asymmetric perturbations introduced on the left and right sides to reduce excessive symmetry. Additionally, a high-speed gradient anomaly is incorporated at the bottom. The final velocity values range from a maximum of \(4.97 \,  \ \text{km/s}\) to a minimum of \(1.87 \,  \ \text{km/s}\).

\begin{figure}
    \centering
    \includegraphics[width=0.8\linewidth]{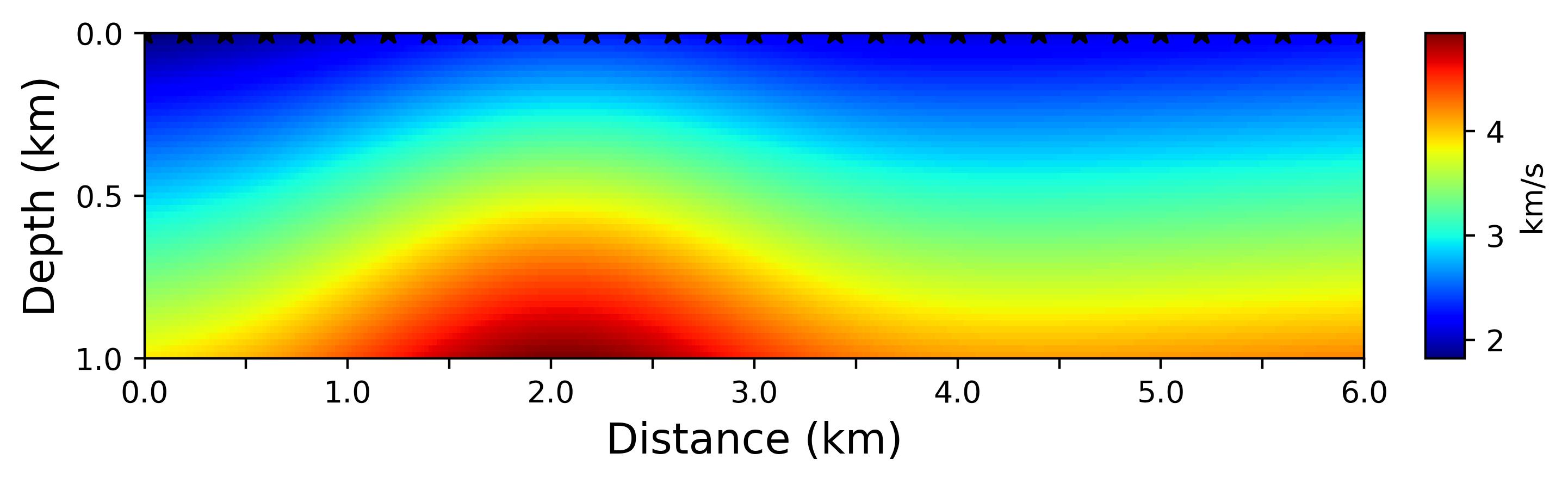}
    \caption{\centering Lens velocity model}
    \label{l_v}
\end{figure}

\begin{figure}
    \centering
    \includegraphics[width=0.8\linewidth]{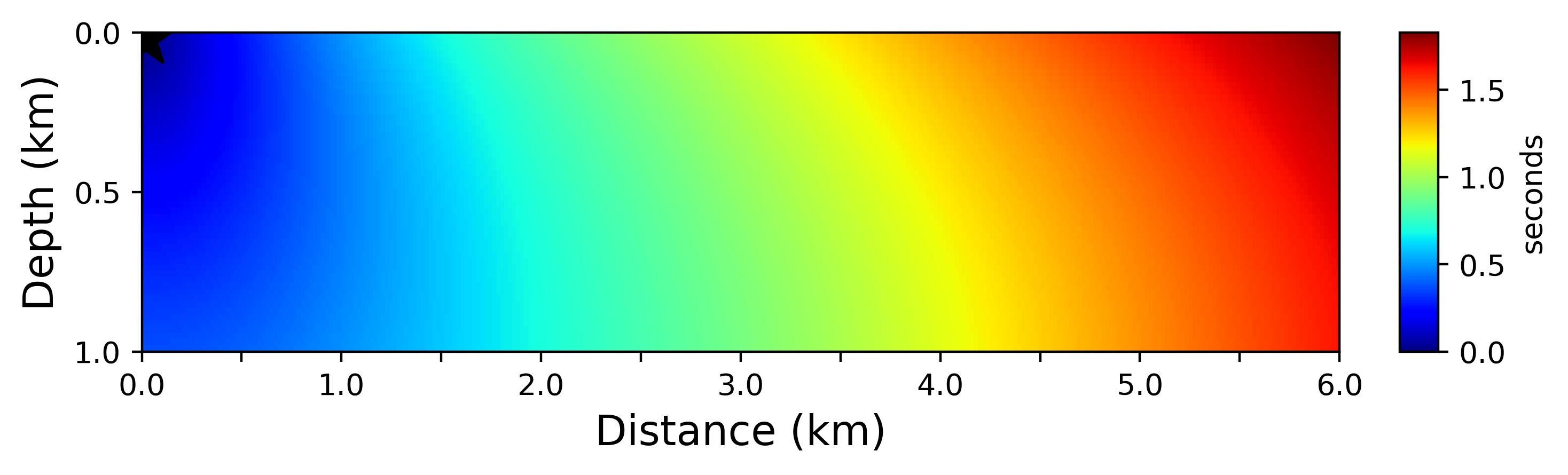}
    \caption{\centering Traveltime of lens velocity model(taking the leftmost source)}
    \label{l_t}
\end{figure}

The neural network used for this model consists of 10 hidden layers, each  layer containing 20 neurons. A total number of 50,000 sets of model coordinates and source coordinates are randomly generated as inputs to the neural network. The number of source and receiver pairs is 435. Both traditional PINN and RC-PINN approaches are trained with their respective loss functions for 10,000 epochs. The velocity  model is shown in Figure~\ref{l_v}. The true traveltime of one source located at (0 km, 0 km) obtained from FMM as the reference solution is illustrated in Figure~\ref{l_t}.

We apply traditional PINN, RC-PINN, and RC-PINN with dynamic weights (RC-PINN-w) to predict traveltimes for the lens velocity model. The same neural network architecture was used for all methods, and sufficient training was conducted. 

\begin{figure}
    \centering
    \includegraphics[width=0.6\linewidth]{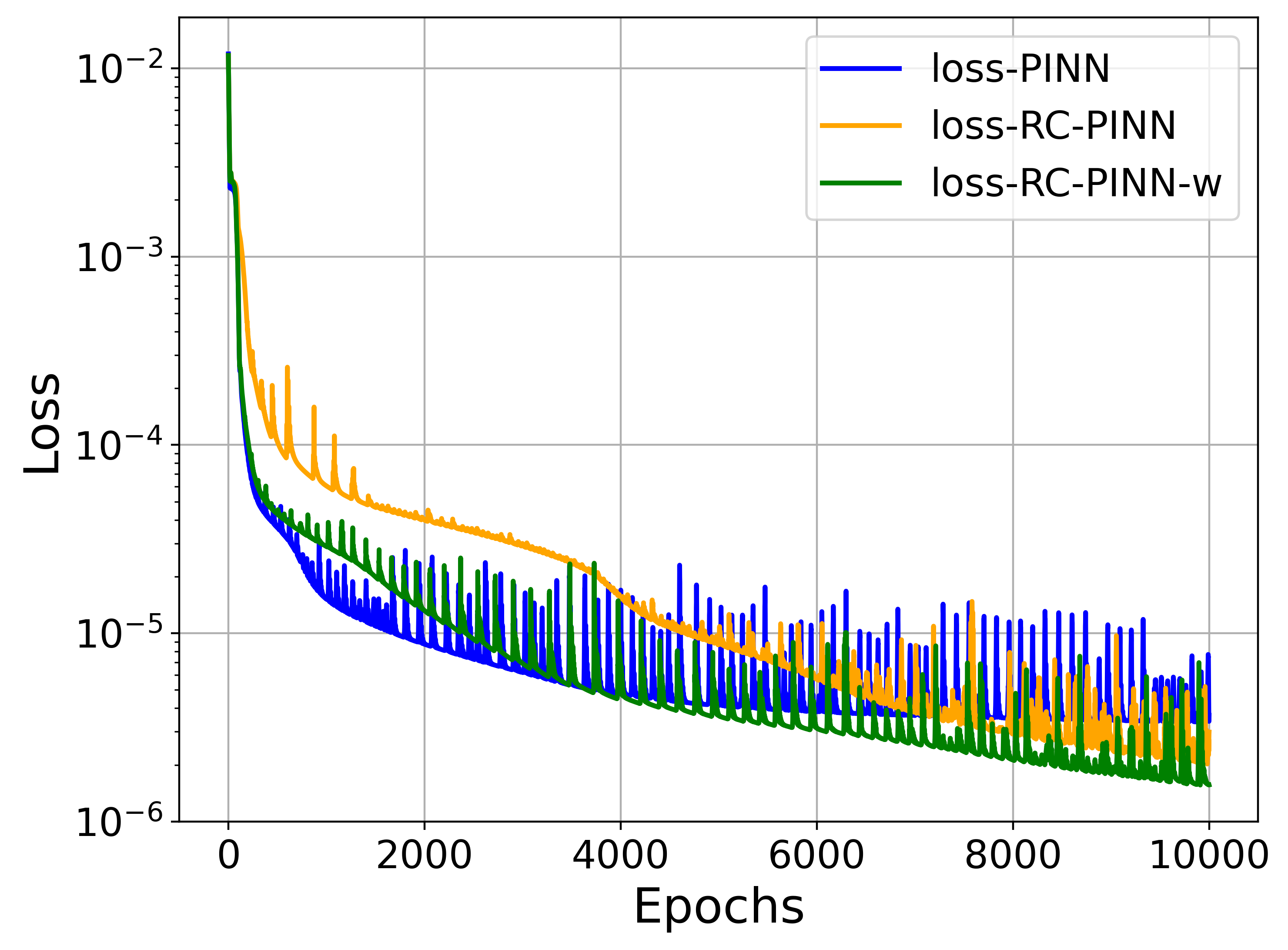}
    \caption{\centering Physical loss terms of PINN, RC-PINN, and RC-PINN-w for of lens velocity model}
    \label{l_l}
\end{figure}

For a detailed comparison and analysis of the three PINN-based methods, we focused on the loss function corresponding to the eikonal equation's physics constraints. Figure~\ref{l_l} illustrates the loss curves with the number of epochs during the training process for the three methods.

As shown in Fig. 8, the losses for all three methods decrease rapidly at the early stage of training. Among the three methods, the traditional PINN method exhibits the fastest loss decrease during the first 4000 epochs, but with significant fluctuations. In contrast, though RC-PINN and RC-PINN-w show a slower initial decrease, their final loss values are notably smaller than those of the traditional PINN. RC-PINN also exhibits some fluctuations; however, the fluctuation range is smaller, indicating improved stability of the network training. Additionally, the introduction of dynamic weights significantly accelerates the loss function's decrease, greatly enhancing traveltime prediction accuracy.

To further compare the three methods, we generate traveltime error maps of PINN, RC-PINN, and RC-PINN-w, with the results shown in Figure 9a-9c. Due to the significant scale differences in the \(x\)- and \(z\)-directions of the model, the neural network used in the traditional PINN may fail to accurately predict traveltimes for locations far from the source. This leads to larger errors on the right bottom corner of the model for the traditional PINN. In contrast, the traveltime error produced by RC-PINN is smaller. Furthermore, after introducing dynamic weights, the error is further reduced, effectively eliminating traveltime errors for locations far from the source.

\begin{figure}
    \centering
    \includegraphics[width=0.8\linewidth]{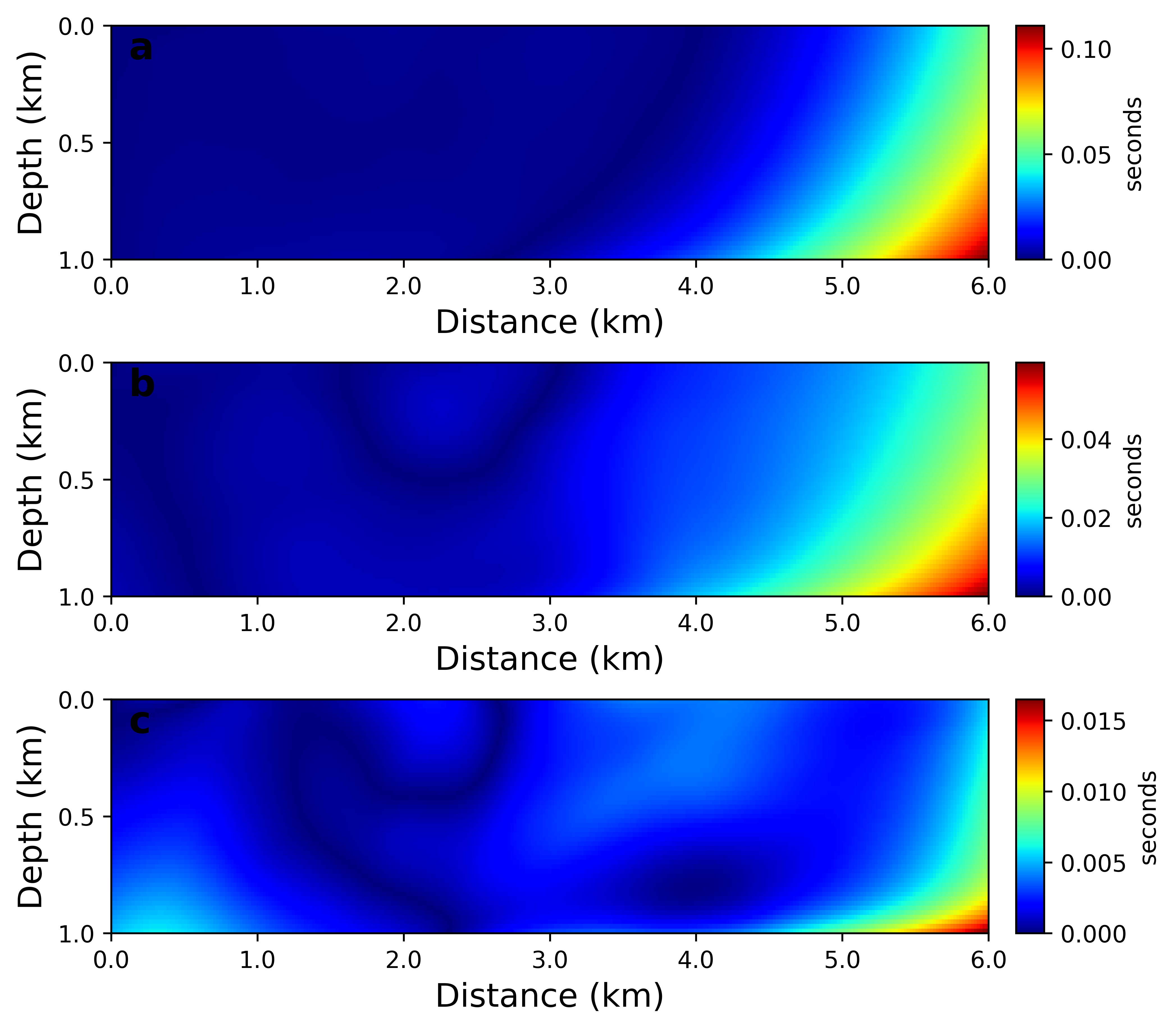}
    \caption{\centering The traveltime errors of lens velocity model using traditional PINN (a), RC-PINN(b) and RC-PINN-w (c).}
    \label{l_t_e}
\end{figure}
To further compare the performance of the three methods, we output their \(L_2\)  error and maximum traveltime error, with the results shown in Table~\ref{2}. As illustrated in Table 2, the \(L_2\)  error and maximum traveltime error of RC-PINN are lower than those of the traditional PINN, demonstrating that RC-PINN provides more accurate results for traveltime prediction. Furthermore, with the introduction of dynamic weights, the training efficiency is improved, resulting in further increases in accuracy under the same number of training epochs.

\begin{table}[htbp]
\centering
\caption{L2 error and maximum error of traveltimes from PINN,  RC-PINN, and RC-PINN-w of Lens velocity model }\label{tab1}%
\begin{tabular}{@{}llll@{}}
\toprule
Method   & L2  Error & Maximum Error (s) \\
\midrule
PINN      & 2.71e-2            &0.125             \\
RC-PINN     & 9.17e-3            &0.057             \\
RC-PINN-w & 2.72e-3            &0.016             \\
\bottomrule
\label{2}
\end{tabular}
\end{table}

\subsection{Overthrust velocity model}\label{subsec2}
\begin{figure}
    \centering
    \includegraphics[width=0.8\linewidth]{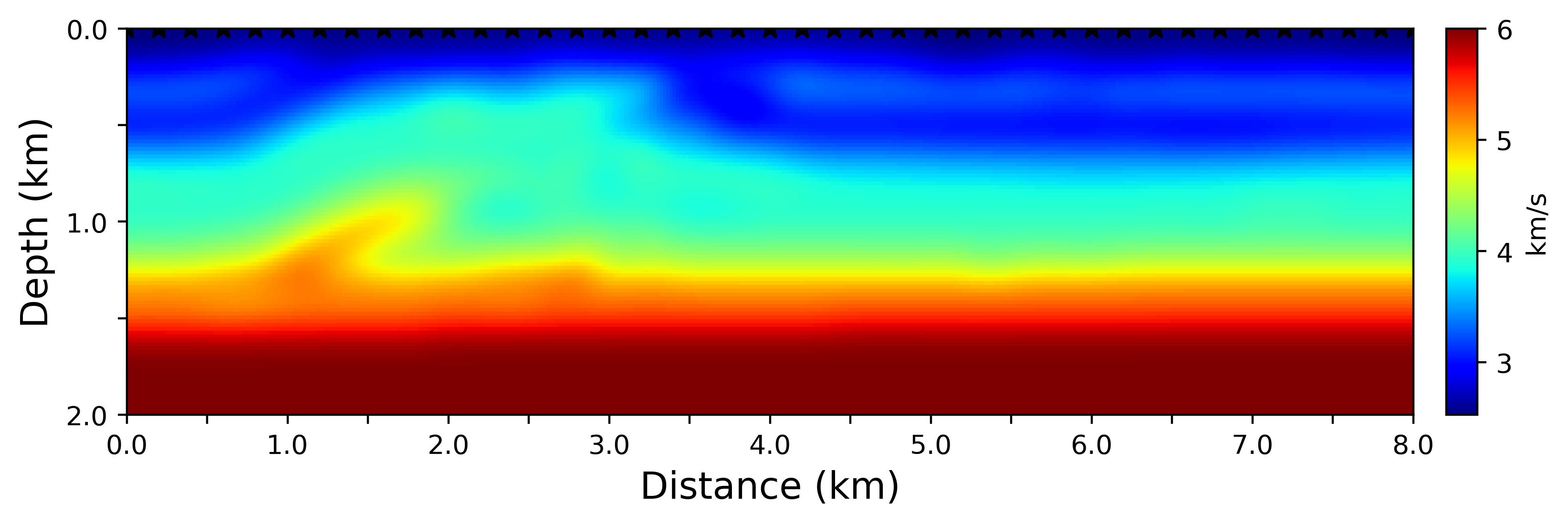}
    \caption{\centering Overthrust velocity model}
    \label{o_v}
\end{figure}

\begin{figure}
    \centering
    \includegraphics[width=0.8\linewidth]{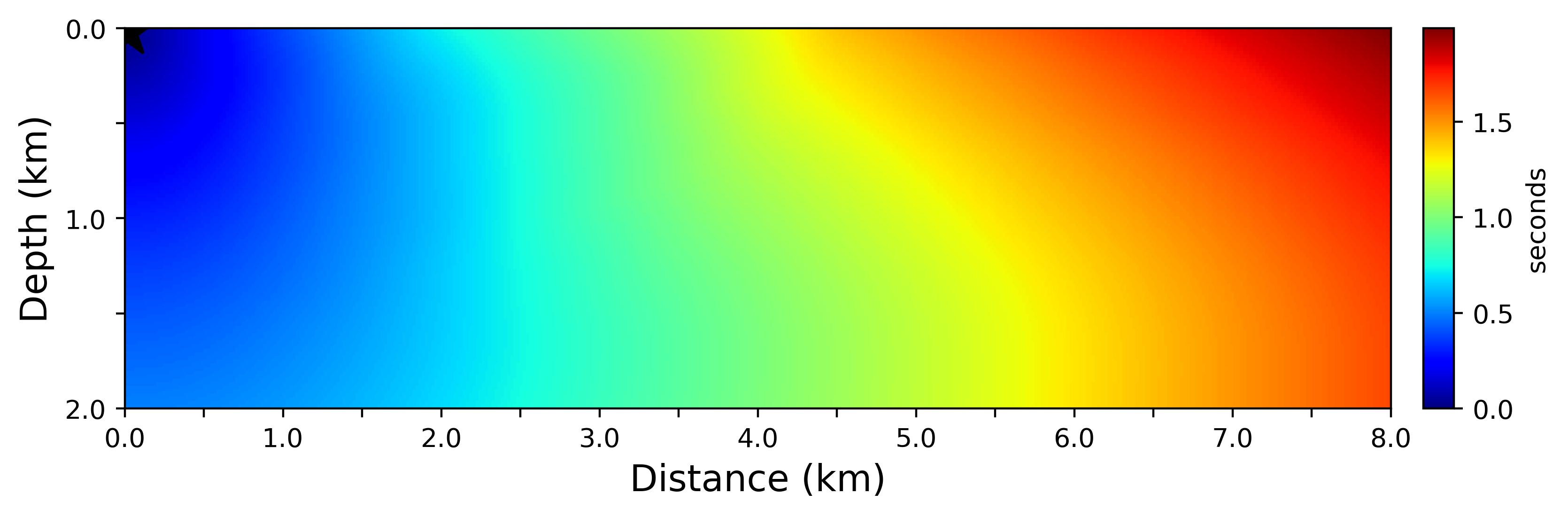}
    \caption{\centering Traveltime of Overthrust velocity model(taking the leftmost source)}
    \label{o_t}
\end{figure}

Next, we apply the Overthrust velocity model for traveltime prediction. The model size is \(401 \times 101\) with \(0.02 \,  \ \text{km}\) apart in both horizontal and vertical directions, covering a region of \(8 \times 2 \,  \ \text{km}^2\). The neural network for this model consists of 10 hidden layers, each with 50 neurons. A total number of 50,000 sets of model coordinates and source coordinates are randomly generated as inputs to the network. The number of source and receiver pairs is 780. Both traditional PINN and RC-PINN approaches are trained with their respective loss functions for 50,000 epochs. The velocity model is shown in Figure~\ref{o_v}. The true traveltime of one source located at (0 km, 0 km) obtained from FMM as the reference solution is illustrated in Figure~\ref{o_t}.

\begin{figure}
    \centering
    \includegraphics[width=0.6\linewidth]{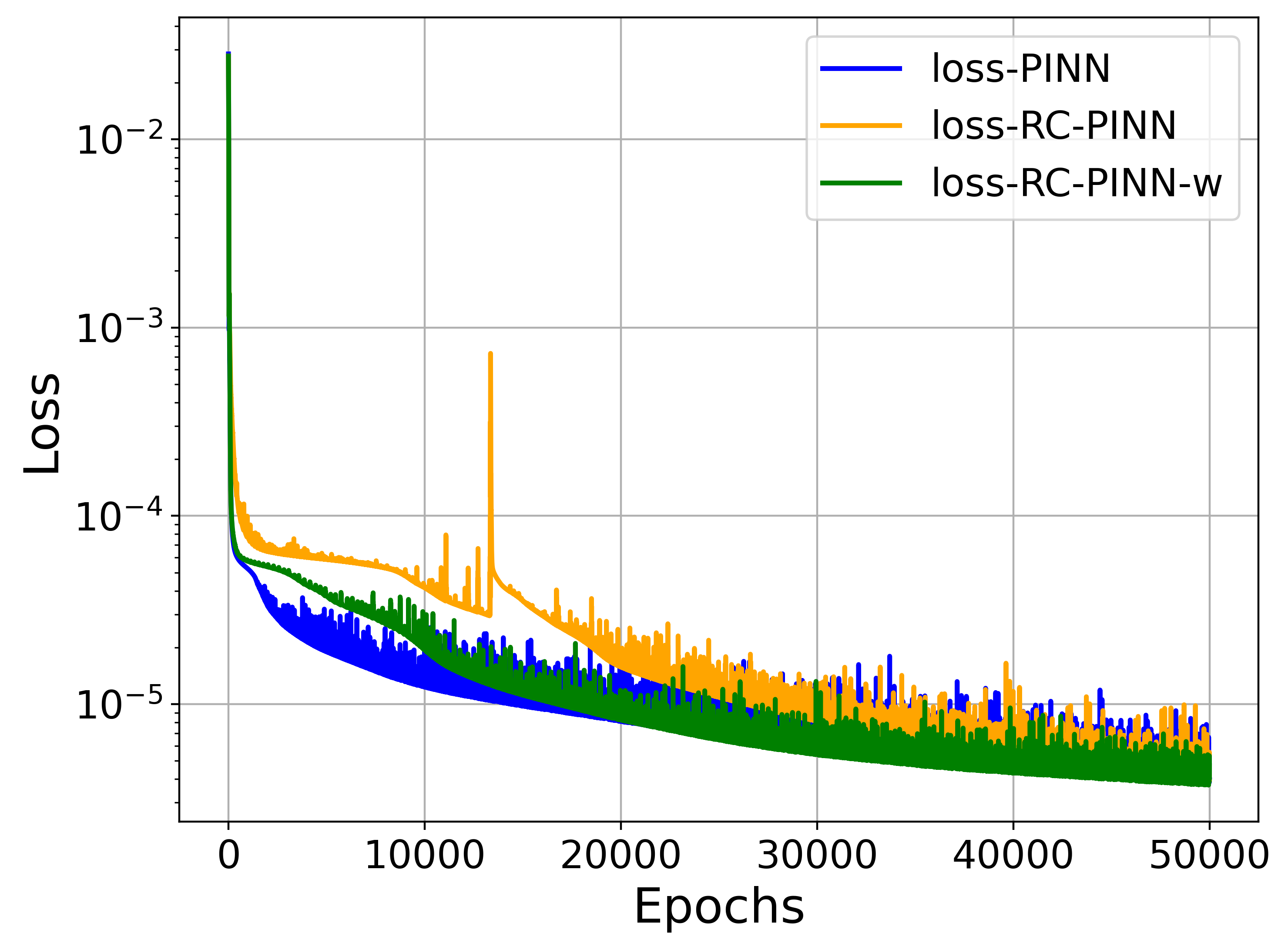}
    \caption{\centering Physical loss terms of PINN, RC-PINN, and RC-PINN-w for Overthrust velocity model}
    \label{o_l}
\end{figure}

We apply traditional PINN, RC-PINN, and RC-PINN-w to predict traveltimes for the Overthrust velocity model. The same neural network architecture was used for all methods, and sufficient training was conducted. Figure~\ref{o_l} illustrates the loss curves with the number of epochs during the training process for the three methods.

As shown in Fig. 12, the losses for all three methods decrease rapidly at the early stage of training. The traditional PINN method exhibits the fastest loss decrease but is eventually surpassed by RC-PINN and RC-PINN-w. Additionally, RC-PINN-w shows a faster decrease compared to RC-PINN with much smaller fluctuations, demonstrating that the introduction of dynamic weights improves both convergence speed and traveltime prediction accuracy.

To further compare the three methods, we generate traveltime error mapsof PINN, RC-PINN, and RC-PINN-w, with the results shown in Figure 13a-13c. The error produced by the traditional PINN method is relatively large in the middle region. It is because the velocity model in middle region is more complex, making it challenging for traditional PINN to accurately predict traveltimes. In contrast, RC-PINN  reduces the error in this region compared to traditional PINN. Furthermore, after the introduction of dynamic weights, the traveltime error in this region is further reduced, indicating that RC-PINN-w demonstrates better performance in complex velocity models.

\begin{figure}
    \centering
    \includegraphics[width=0.8\linewidth]{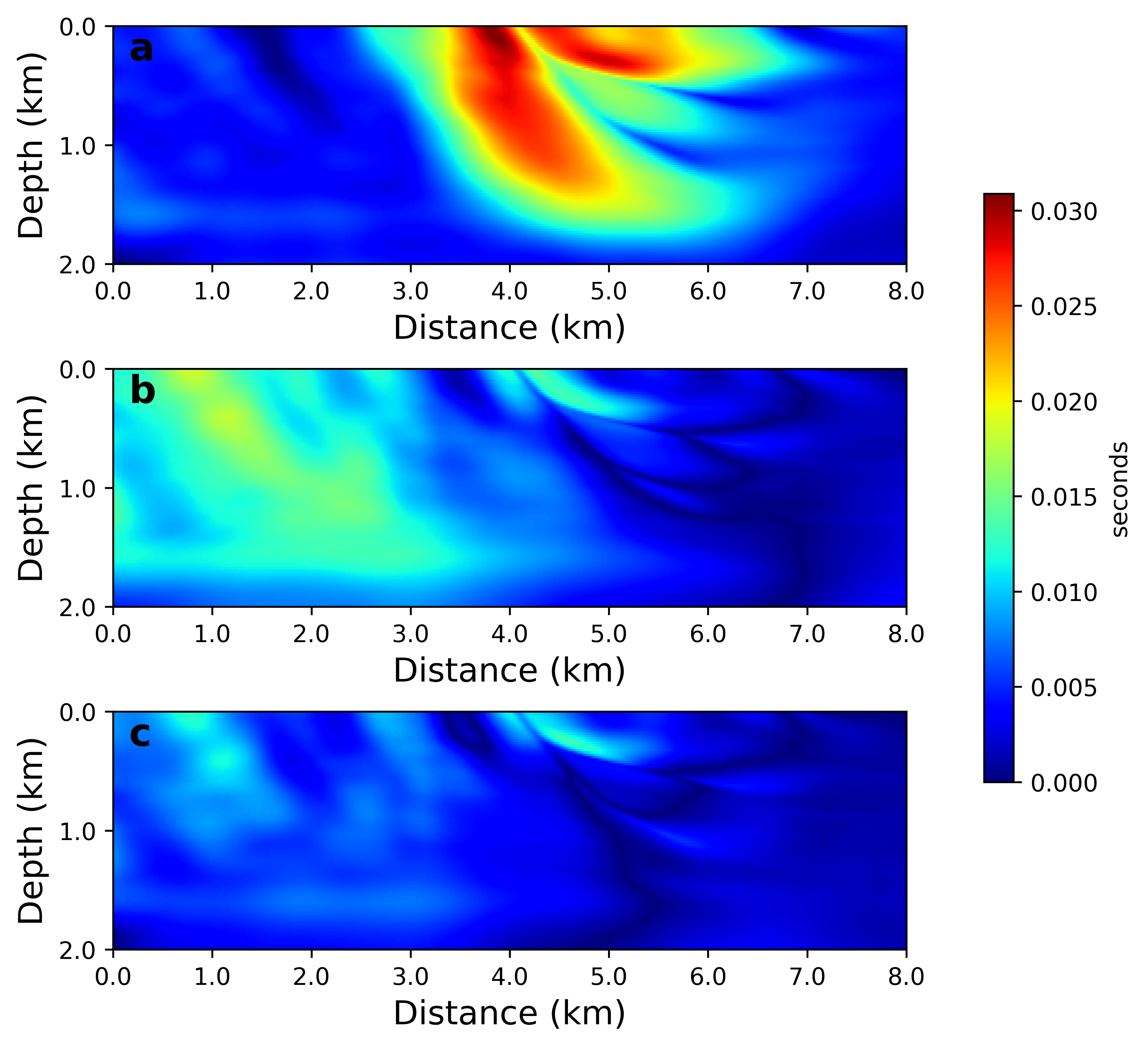}
    \caption{\centering The traveltime errors of Overthrust velocity model using traditional PINN (a), RC-PINN(b) and RC-PINN-w (c).}
    \label{o_t_e}
\end{figure}

To further compare the performance of the three methods, we output their \(L_2\)  error and maximum traveltime error, with the results shown in Table~\ref{3}. As illustrated in Table 3, the \(L_2\)  error and maximum traveltime error of RC-PINN are lower than those of traditional PINN. After incorporating dynamic weights, the training efficiency is improved, resulting in further increases in accuracy under the same number of training epochs.

\begin{table}[htbp]
\centering
\caption{L2 error and maximum error of traveltimes from PINN,  RC-PINN, and RC-PINN-w of Overthrust velocity model }\label{tab1}%
\begin{tabular}{@{}llll@{}}
\toprule
Method   & L2  Error & Maximum Error (s) \\
\midrule
PINN     & 8.91e-3            &0.025             \\
RC-PINN     & 6.71e-3            &0.017            \\
RC-PINN-w & 5.52e-3            &0.013             \\
\bottomrule
\label{3}
\end{tabular}
\end{table}

\subsection{3D Velocity Model}\label{subsec2}
\begin{figure}[htbp]
    \centering
    \begin{minipage}[b]{0.48\textwidth}
        \centering
        \includegraphics[width=\textwidth]{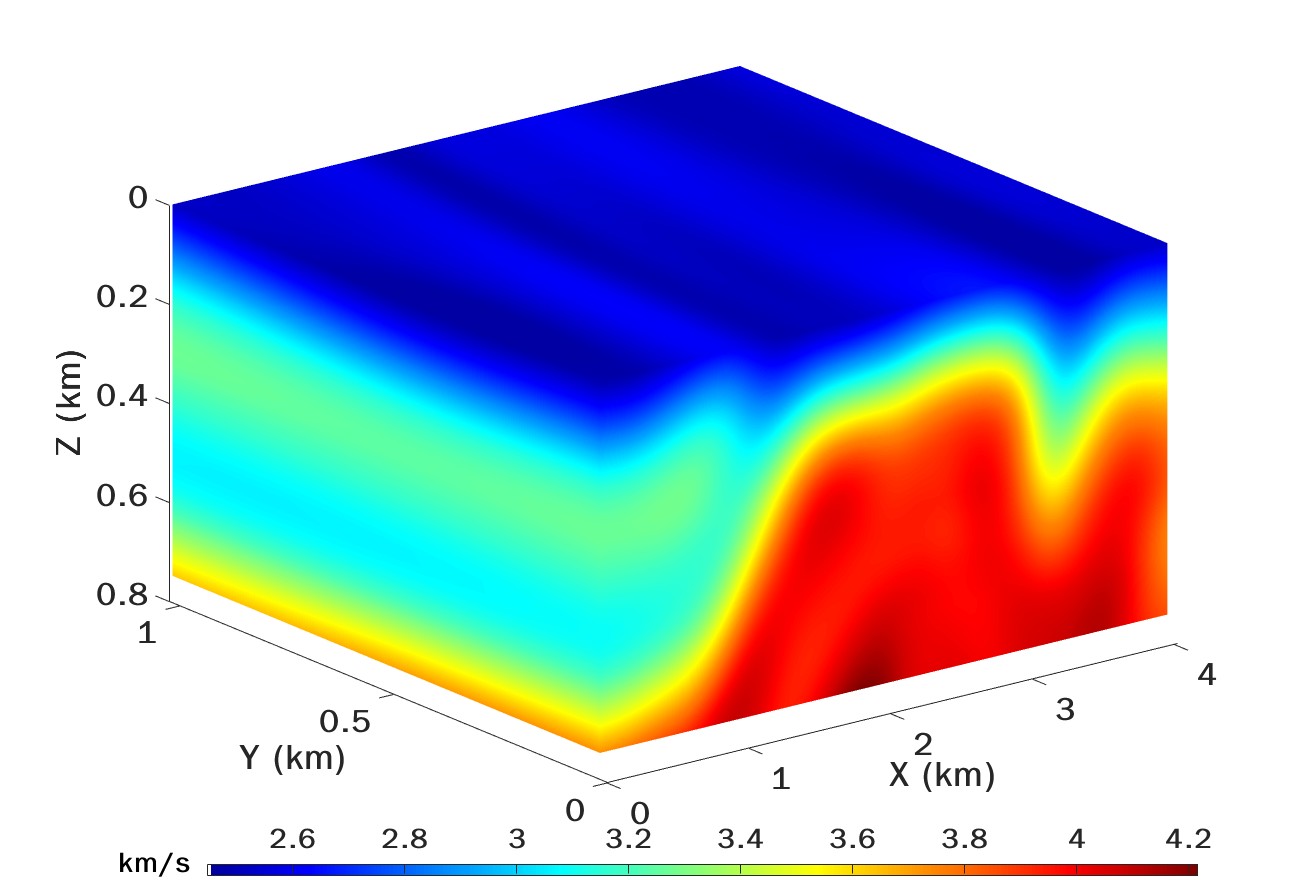}
        \caption{\centering 3D velocity model}
        \label{3_v}
    \end{minipage}
    \hfill
    \begin{minipage}[b]{0.48\textwidth}
        \centering
        \includegraphics[width=\textwidth]{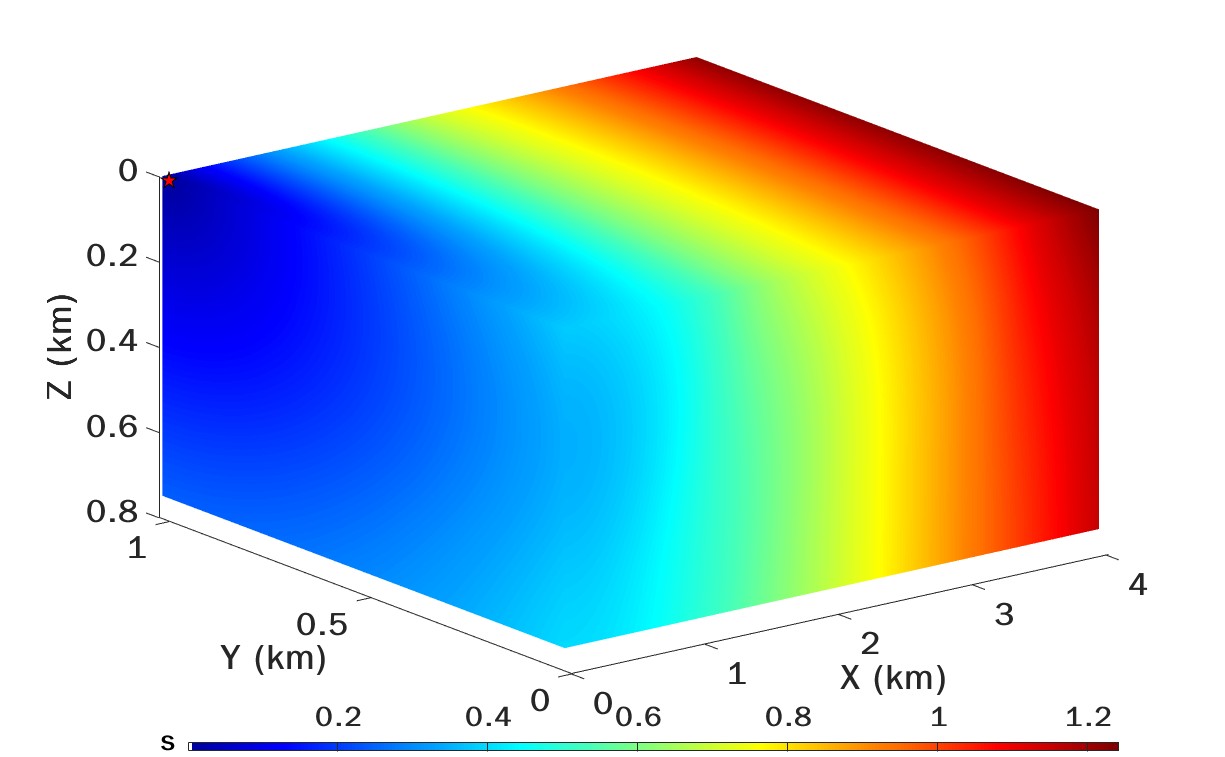}
        \caption{\centering Traveltime of 3D velocity model}
        \label{3_t}
    \end{minipage}
\end{figure}

To verify the applicability of RC-PINN in 3D velocity models, we apply a 3D velocity model  for traveltime simulation. The model has dimensions \(31 \times 161 \times 41\), spaced \(0.025 \, \ \mathrm{km}\) apart in three mutually orthotropic directions, covering a region of \(0.75 \times 4 \times 1 \,  \ \mathrm{km^3}\). The neural network used for this model consists of 10 hidden layers, each  layer containing 32 neurons. A total number of 100,000 sets of model coordinates and source coordinates were randomly generated as inputs to the neural network. The number of source and receiver pairs is 3570.  The neural network was then trained using both the traditional PINN and RC-PINN methods, with their respective loss functions for 20,000 epochs. The velocity model is shown in Figure~\ref{3_v}.  The true travel time of one source located at (0 km, 0 km, 1 km) obtained from FMM as the reference solution is illustrated in Figure~\ref{3_t}.

\begin{figure}
    \centering
    \includegraphics[width=0.6\linewidth]{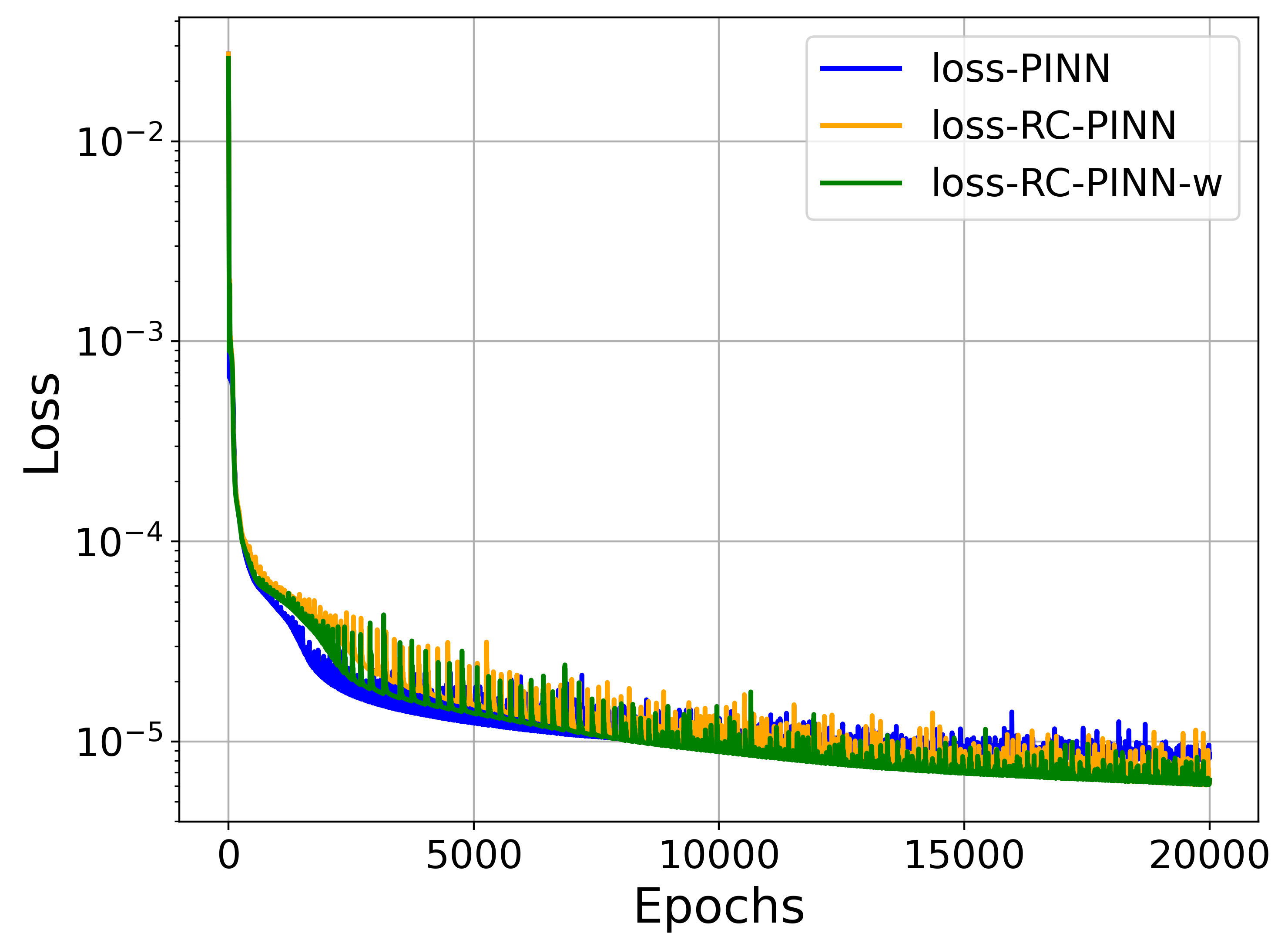}
    \caption{\centering Physical loss terms of PINN, RC-PINN, and RC-PINN-w for 3D velocity model}
    \label{3_l}
\end{figure}

The loss curves with the number of epochs during the training process for traditional PINN, RC-PINN, and RC-PINN-w to predict traveltimes are shown in Figure~\ref{3_l}. We get the same conclusion that RC-PINN exhibits higher accuracy than traditional PINN and RC-PINN-w accelerates the loss function's decrease.

To further compare the three methods, we generate traveltime error maps of PINN, RC-PINN, and RC-PINN-w, with the results shown in Figure 17a-17f. It can be observed that the neural network used in the traditional PINN exhibits larger errors, particularly in the region surrounding (0 km, 0 km, 0 km). RC-PINN reduces the error in this region, but still shows higher errors near the middle of the upper surface. In contrast, RC-PINN-w  reduces the traveltime error in these regions. 

To further compare the performance of the three methods, we output their \(L_2\) error and maximum traveltime error, with the results summarized in Table~\ref{4}. As illustrated in Table 4, we get the same conclusions as we get in the 2d model that  the \(L_2\)  error and maximum traveltime error of RC-PINN are lower than those of the traditional PINN and that RC-PINN-w improves the training efficiency, resulting in further increases in accuracy under the same number of training epochs.


\begin{figure}[htbp]
    \centering
    \begin{subfigure}{0.45\textwidth}
        \centering
        \includegraphics[width=\linewidth]{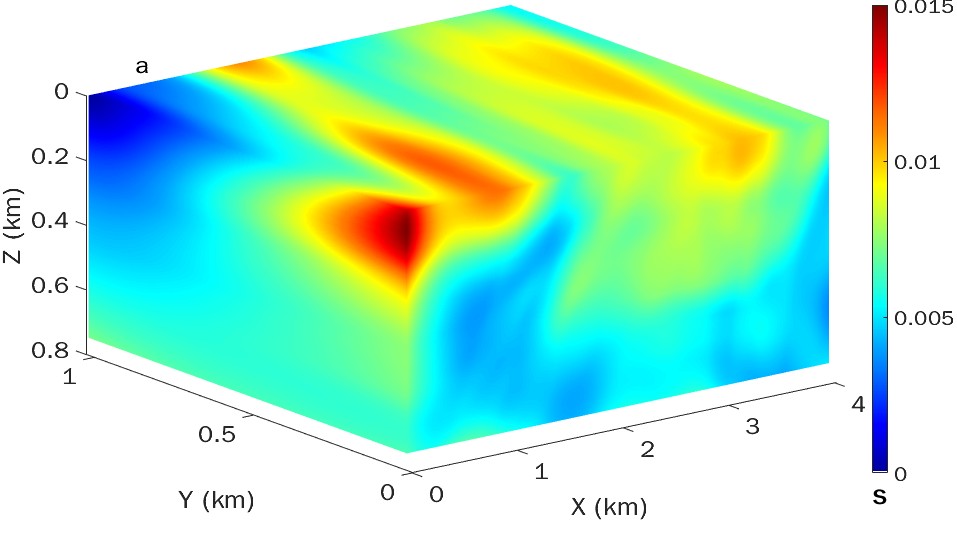}
    \end{subfigure}
    \begin{subfigure}{0.45\textwidth}
        \centering
        \includegraphics[width=\linewidth]{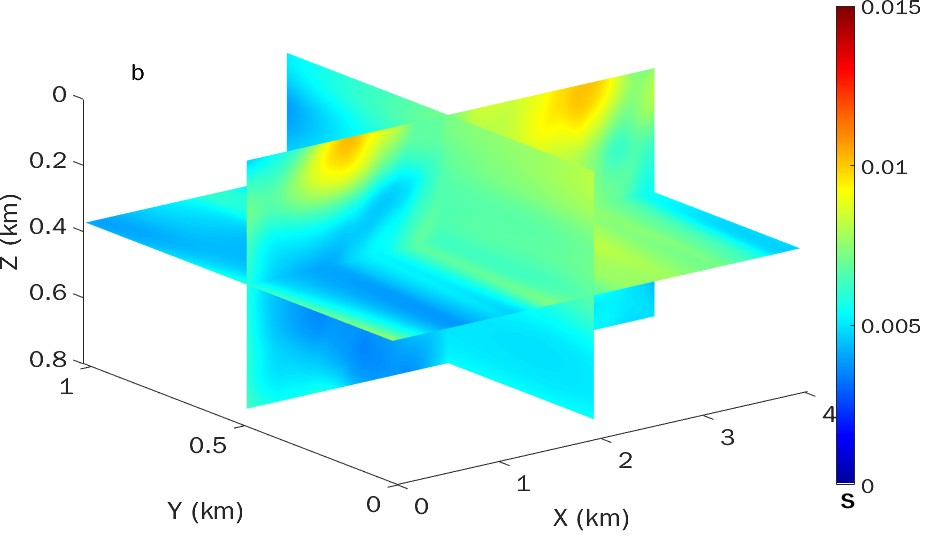}
    \end{subfigure}

    \begin{subfigure}{0.45\textwidth}
        \centering
        \includegraphics[width=\linewidth]{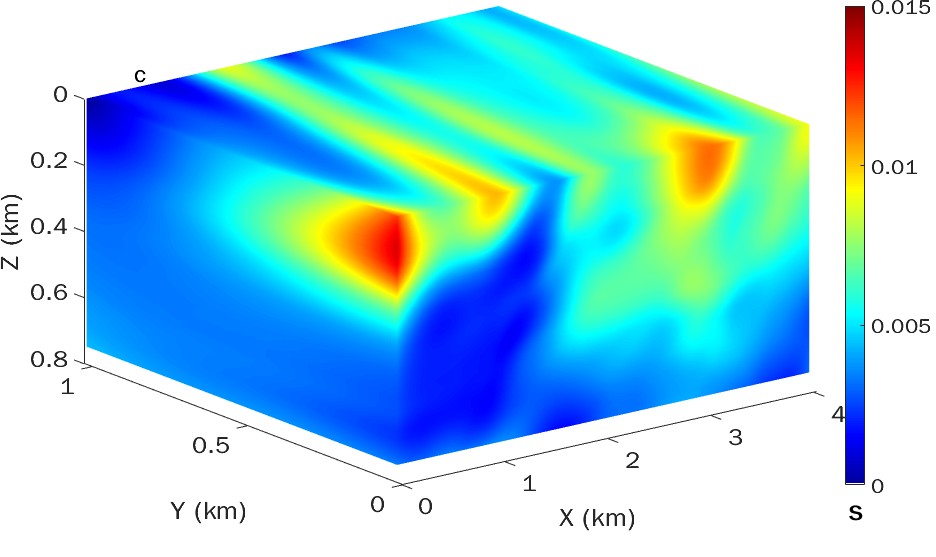}
    \end{subfigure}
    \begin{subfigure}{0.45\textwidth}
        \centering
        \includegraphics[width=\linewidth]{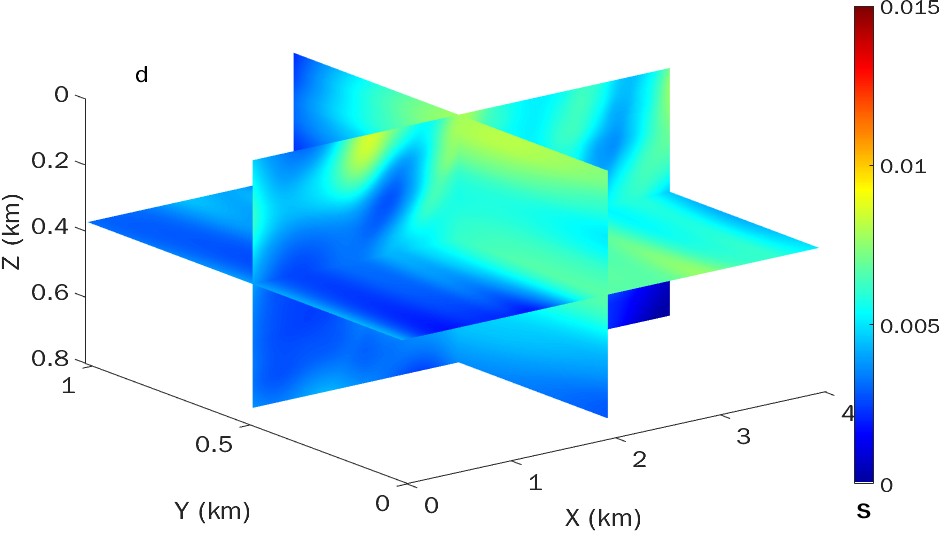}
    \end{subfigure}

    \begin{subfigure}{0.45\textwidth}
        \centering
        \includegraphics[width=\linewidth]{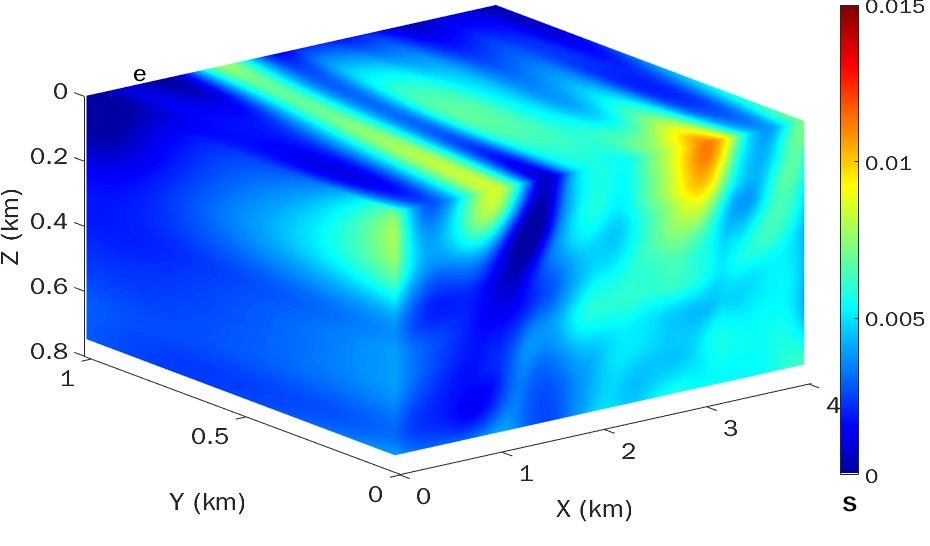}
    \end{subfigure}
    \begin{subfigure}{0.45\textwidth}
        \centering
        \includegraphics[width=\linewidth]{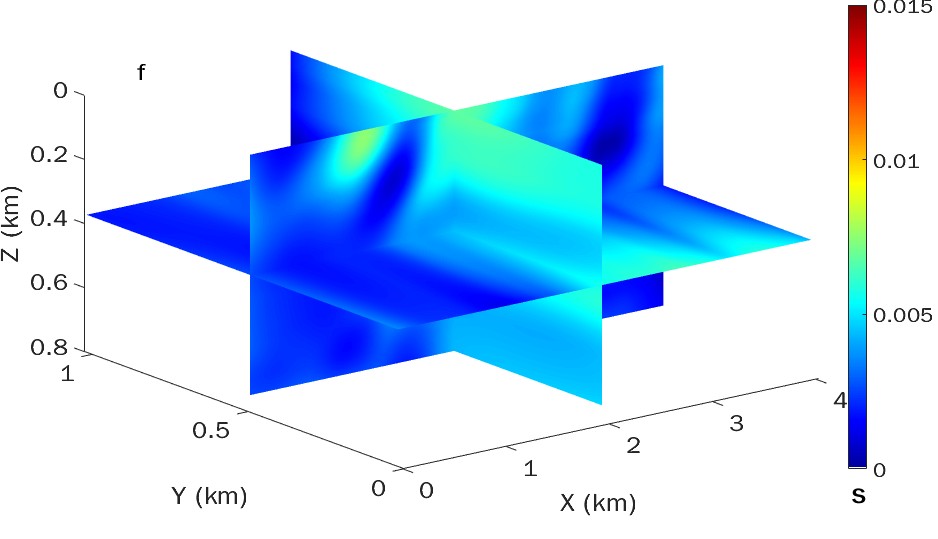}
    \end{subfigure}

    \caption{\centering The traveltime errors of 3D velocity model using PINN(a,b),RC-PINN(c,d)and RC-PINN-w(e,f).}
    \label{3_l_e}
\end{figure}

\begin{table}[htbp]
\centering
\caption{L2 error and maximum error of traveltimes from PINN,  RC-PINN, and RC-PINN-w of 3D velocity model }\label{tab1}%
\begin{tabular}{@{}llll@{}}
\toprule
Method   & L2  Error & Maximum Error (s) \\
\midrule
PINN      & 7.62e-3            &0.0150             \\
RC-PINN     & 7.11e-3            &0.0134            \\
RC-PINN-w & 6.73e-3            &0.0117             \\
\bottomrule
\label{4}
\end{tabular}
\end{table}

\section{Conclusion}\label{sec3}
We  proposed a reciprocity-constrained physics-informed neural network (RC-PINN) method to predict seismic first-arrival traveltimes. By incorporating the principle of reciprocity into the PINN-based eikonal equation solver, without additional training costs, this approach can effectively reduce traveltime simulation errors , especially for velocity models with large directional scale differences. Additionally, a dynamic weighting strategy is introduced between  the physical loss  constrained by the eikonal equation and the reciprocity-constrained loss, adapting with the number of training epochs. This adjustment greatly improve training convergence performance. The effectiveness and superiority of RC-PINN have been validated using multiple 2D and 3D velocity models. 

\section*{Acknowledgements}
This research is funded by the National Key Research and Development Program of China
(Grant No.2023YFC3707901), and partially by the Key Program of the National Natural
Science Foundation of China (Grant No.42430801).
\bibliography{cas-refs}

\end{document}